\documentclass[preprint,preprintnumbers,amsmath,amssymb]{revtex4}

\usepackage{graphicx}
\usepackage{dcolumn}
\usepackage{bm}

\usepackage{axodraw}
\usepackage{epsfig}

\begin{document}

\title{Direct Photon Production in Association With A Heavy Quark At Hadron 
Colliders}

\author{T. Stavreva}

\author{J.F. Owens}

\affiliation{Department of Physics, Florida State University,\\
             Tallahassee, FL 32306-4350,USA}

\begin{abstract}
Results of a next-to-leading order calculation of the inclusive cross section 
for a photon and a heavy quark (charm or bottom), 
$p \bar p / pp \rightarrow \gamma +Q +X$, are presented.  Pointlike photon 
subprocesses through ${\cal O}(\alpha\alpha_{s}^2)$ and fragmentation 
subprocesses through ${\cal O}(\alpha_{s}^3)$ are included.  The calculation 
is performed using a phase space slicing technique so that the effects of 
experimental cuts can be included.  Results for the ratios of the charm and 
bottom cross sections are presented and the systematics of the various 
subprocesses for both the Tevatron and the LHC are compared and contrasted. 
\end{abstract}

\bibliographystyle{unsrt} 

\maketitle
\newpage
\setcounter{page}{1}
\section{Introduction}

Large momentum transfer processes have a long history of providing information 
on the substructure of hadrons, the nature of their constituents, and how they 
interact\cite{JO:review}. Photons provide an excellent probe for such purposes 
due to their pointlike electromagnetic coupling to the quark constituents of 
the colliding hadrons. The production of large momentum transfer photons 
has played dual roles of providing information of parton distribution 
functions (PDFs) \cite{ABFOW} and testing the adequacy of the perturbative 
techniques used to calculate the hard scattering subprocesses 
\cite{Huston:gamma, Aurenche:old_study, Aurenche:new_study}. 

The single photon cross section, either inclusive or subject to photon 
isolation cuts, provides the basic observable for direct photon studies.  
The calculation of this cross section involves integrations over the phase 
space variables of the accompanying partons, thereby limiting the information 
which can be obtained about the underlying subprocesses.  More information 
can be obtained if one can measure an associated jet as has been done 
recently by the D\O collaboration \cite{Dzero:photonjet}.  Even here, 
however, one is summing over many subprocesses with various flavors of 
partons.  Additional information can be obtained if the flavor of the 
produced jets is tagged.  Identifying jets which contain a heavy quark 
provides exactly this opportunity.

In this paper we investigate in detail one particular piece of the direct 
photon calculation, namely the associated production of direct photons and 
heavy quarks, where the heavy quarks are either charm or bottom.  Some of the 
contributing subprocesses are dependent on the charge of the heavy quark 
while others are not.  By considering both charm and bottom quarks one can 
examine the relative roles of the two contributions.  In some kinematic 
regions the results are dependent on the heavy quark PDFs, opening the 
possibility of testing the current calculation of such PDFs.  

New measurements of this process by the D\O\  and CDF groups are in progress 
and should be available in the near future.  A comparison with these 
measurements will be presented in a forthcoming paper.

The production of heavy quarks at high-$p_T$ has the potential to generate 
logarithms of the form $ln(p_T/m_Q)$ as a result of collinear configurations 
involving $Q\rightarrow Qg$ and $g\rightarrow Q \bar Q$.  These logarithms can 
be resummed via the DGLAP equations for appropriately defined PDFs and 
fragmentation functions (FFs).  This is commonly referred to as a variable 
flavor scheme with either four or five flavors.  In such schemes the heavy 
quarks are treated as massless.  The dominant remaining mass effect is due to 
the imposition of a threshold constraint such that the PDFs and FFs are taken 
to be zero when the hard scattering scales are smaller than the quark mass.  
The calculation presented here has been performed using the variable flavor 
scheme. 
  
There have been previous studies of this process 
\cite{Bailey:charm, Berger:analyt,Vogel:mass,Berger:spin}.  In Ref.
\cite{Bailey:charm, Berger:analyt} results are shown for the production of a 
direct photon plus charm.  Here we also provide results for direct photon plus 
bottom production, and a comparison between the charm and bottom case.  
As noted above, this comparison depends on the relative roles of terms which 
depend on the heavy quark charge and those which do not.  We also extend the 
calculation by treating the photon fragmentation contribution to 
next-to-leading-order (NLO). The previously cited references treated this 
component only in leading order (LO). One further technical point is related 
to the treatment of final state collinear singularities in the case when a 
gluon is emitted collinearly to a final state heavy quark.  
In Ref.\cite{Bailey:charm} this singularity is factored into a charm FF. The 
present calculation is for the case of a photon produced in association with a 
jet which has been tagged as containing a heavy quark, so the fragmentation 
function is replaced by an appropriate jet definition.

In order to be able to work in the massless approximation the heavy quarks and 
photons produced need to carry a transverse momentum, $p_T$, which is few 
times larger than the mass of the heavy quark $m_Q$, i.e. 
$p_T \geq 10 {\rm\ GeV}$.  Since the lower bounds for the values of the 
transverse momenta for direct photons and heavy quarks measurable at both 
the D\O\  and CDF collaborations at Fermilab are above 
$p_T \geq 10 {\rm\ GeV}$, a comparison with a massless calculation is 
appropriate.  If however we are close to the threshold region for production 
of heavy quarks, {\itshape i.e.} $m_Q\sim p_T$, their mass needs to be 
retained, and they are assumed to be only produced externally in pairs, 
as end products of the hard-scattering.  This is called the Fixed Flavor 
Number scheme (FFN), as the number of flavors that compose the nucleon 
remains fixed and it does not depend on the center of mass energy of the hard 
scattering.  Here the proton is assumed to be composed only of light flavors, 
and in lowest order there are only two subprocesses in which the direct 
photon and heavy quarks can be produced. These are 
$gg\rightarrow \gamma Q\bar Q$, and $q\bar q\rightarrow \gamma Q\bar Q$.  A 
study of the case when the quark masses are retained has been done at LO 
\cite{Vogel:mass}, where a comparison between the LO massive and massless 
approaches has been shown to give very similar results.  In Ref.
\cite{Vogel:mass} a differential cross section in the transverse momentum of 
the photon up to values of $p_{T\gamma} \sim 50 {\rm\ GeV}$ is presented.  
There the difference between the two approaches in the LO is minimal. 

This paper proceeds as follows: in Section II a description of the theory and 
techniques for the calculation are outlined.  In Section III results for the 
differential cross sections are shown.  Predictions for both the Tevatron and 
LHC are presented and compared.  The effects of including the NLO 
fragmentation terms are shown, as well as the effect of the use of different 
charm PDFs on the cross section.  In Section IV we summarize our findings.

\section{Theory}

\begin{figure}[h]
\begin{center}
\includegraphics[scale=1.0,angle=0]{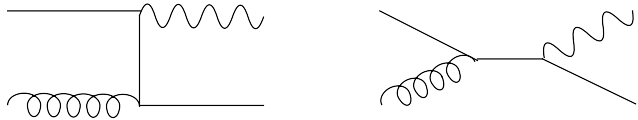}
\end{center}
\caption{\label{Compton} \small{Compton Scattering}}
\end{figure}

\begin{figure}[h]
\begin{center}
\includegraphics[scale=1.0,angle=0]{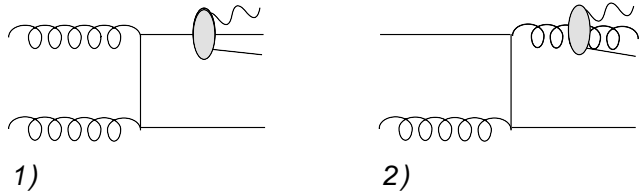}
\end{center}
\caption{\label{Fragm_LO} \small{An example of Leading Order Fragmentation 
Contributions 1) $gg \rightarrow Q\bar Q \gamma$, where the photon can 
fragment off from either one of the final state heavy quarks, 
2) $gQ\rightarrow gQ \gamma$, where again the photon can fragment off from 
either one of the final state partons, the gluon or the heavy quark}}
\end{figure}

Denoting the electromagnetic and strong couplings by $\alpha \mbox{\rm \ and } 
\alpha_s$, respectively, the lowest order subprocess for the production of a 
photon plus a heavy quark is the Compton subprocess, 
$g+Q\rightarrow \gamma +Q$, shown in Fig.\ref{Compton}. This subprocess is of order 
$\alpha \alpha_s$ and in the variable flavor scheme employed here there is 
only this one subprocess to this order. When one considers higher order 
subprocesses such as $qQ \rightarrow qQ\gamma $, for example, there will be 
a region of phase space where the photon may be emitted collinear with  
the final state $q$, giving rise to a collinear singularity. This singular 
contribution may be absorbed into a photon fragmentation function 
$D_{\gamma/q}$. The photon fragmentation functions satisfy a set of 
inhomogeneous DGLAP equations, the solutions of which are of order 
$\alpha/\alpha_s$. More specifically, one has 

\begin{eqnarray*}
\frac{d D_{\gamma/q}(z,t)}{dt} &=& \frac{\alpha}{2\pi}P_{\gamma/q}(z) +
\frac{\alpha_s}{2 \pi}[D_{\gamma/q}\otimes P_{qq} + D_{\gamma/g}\otimes P_{gq}] \\
\frac{d D_{\gamma/g}(z,t)}{dt} &=& \frac{\alpha_s}{2 \pi}[D_{\gamma/q}\otimes 
P_{qg} + D_{\gamma/g}\otimes P_{gg}] 
\end{eqnarray*}
where $t=ln(Q^2/\Lambda_{QCD}^2)$ and $\otimes$ denotes a convolution.  
Writing $\alpha_s(t)=1/bt$ it is easy to see that the solutions for both 
$D_{\gamma/q} \mbox{\rm \ and } D_{\gamma/g}$ are proportional to both $t 
\mbox{\rm \ and } \alpha$ so that the fragmentation functions may be thought 
of as being \cal{O}($\alpha/\alpha_s$). Therefore, another class of 
contributions of order $\alpha \alpha_s$ consists of $2\rightarrow 2$ 
QCD subprocesses with at least one  heavy quark in the final state convoluted 
with the appropriate photon FF. An example is shown in Fig.\ref{Fragm_LO}.

\begin{figure}[h]
\begin{center}
\includegraphics[scale=1.0,angle=0]{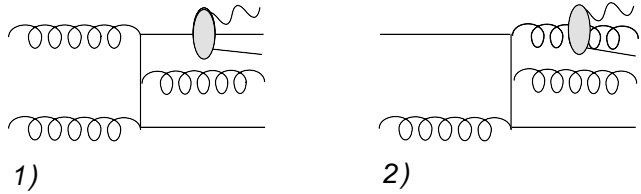}
\end{center}
\caption{\label{Fragm_NLO} \small {An example of Next-to-Leading Order 
Fragmentation Contributions 1) $gg\rightarrow Q\bar Q g \gamma$, where 
again the photon is produced by fragmenting from either of the final state 
partons produced in the hard scattering, 2) another example of NLO 
fragmentation $gQ\rightarrow ggQ \gamma$}}
\end{figure}

At the next order in perturbation theory, $\alpha \alpha_s^2$, the phase space 
for producing a photon in association with a heavy quark increases and now 
there are seven possible subprocesses, which are listed in Table 1.  As in the 
LO case, there are fragmentation contributions that need to be taken into 
account in order to have a complete NLO calculation.  Thus all 
$2\rightarrow 3$ QCD subprocesses of order $\alpha_s^3$ once convoluted 
with $D_{\gamma/q,g}(z,Q)$, give something of the 
NLO: $O(\alpha_s^3)\otimes D_{\gamma/q,g}\sim\alpha_s^3\alpha/\alpha_s=
\alpha\alpha_s^2$, Fig.\ref{Fragm_NLO}.   
\begin{table}[t]
\begin{center}
\begin{tabular}{|c|} \hline
 \em NLO subprocesses \\\hline
$gg\rightarrow \gamma Q\bar Q$\\
$gQ\rightarrow \gamma gQ$\\
$Qq\rightarrow \gamma qQ$ \\
$Q\bar q\rightarrow \gamma qQ$ \\
$q\bar q\rightarrow \gamma Q\bar Q$\\
$Q\bar Q\rightarrow \gamma Q\bar Q$\\
$QQ\rightarrow \gamma QQ$ \\\hline
\end{tabular}
\caption{\small{a list of all $2\rightarrow 3$ NLO hard-scattering 
subprocesses}}
\end{center}
\end{table}

As soon as we go beyond LO, ultraviolet (UV), soft, and collinear divergences 
appear.  The UV singularities arise from virtual diagrams when the momenta of 
the virtual particles go to infinity.  To take care of these the theory is 
renormalized, and the singularities are absorbed into the now renormalized 
strong coupling $\alpha_s$.  The soft singularities appear in the case where 
the energy of a massless particle like the gluon goes to zero, and the 
collinear singularities arise when two massless particles are emitted 
collinearly.  Since generally the energies that are considered are much larger 
than those of the quark masses, the quarks are treated as massless, and the 
calculation is done in the massless approximation.  To take care of the 
divergences the calculation is regularized.  The regularization scheme that is 
used here is Dimensional Regularization (DR).  In DR the scattering amplitudes 
are computed in $d=4-\epsilon$ dimensions, and the singularities are exposed 
as poles in $1/\epsilon$.  These poles cancel, once the virtual, soft and 
collinear contributions are added or they have to be absorbed into Parton 
Distribution Functions (PDFs) and FFs with the use of the DGLAP evolution 
equations.  Once this is done it is safe to go back to $4$ dimensions.

To perform the NLO calculation the two cutoff phase space slicing method 
\cite{Owens:PS} 
is used.  In it the phase space is divided into $2\rightarrow2$
and $2\rightarrow3$ body contributions.  The $2\rightarrow3$ body phase space 
is further divided into a hard region, where no singularities are present, a 
collinear region, where the collinear singularities are present and a soft 
region, where the soft singularities occur.  The separation between the 
different regions is done with the help of two parameters : the soft  
cutoff - $\delta_s$, and the collinear cutoff - $\delta_c$.  In the phase 
space slicing method a gluon is considered to be soft, if 
$E_g<\delta_s\sqrt{\hat s}/2$, where $\sqrt{\hat s}$, is the hard scattering 
center of mass energy.  In order to simplify the integration in this region 
the double pole or soft gluon approximation is used and the 4-momentum of the 
gluon is set to zero, when it appears in the numerator. The collinear region 
is taken to be where either $s_{ij}$ or $|t_{ij}|<\delta_c\hat s$, where 
$s_{ij}, t_{ij}$ are the Mandelstam variables.  In the collinear region the 
leading pole approximation is used, so the relative transverse momentum of the 
two collinear particles in the numerators of the expansion is neglected.  The 
integration over phase space is done with the use of Monte Carlo.  This is 
very useful, as the cross section can be calculated as differential in any 
variable that is needed, such as the transverse momentum - $p_T$ or the 
rapidity - $y$, etc. of a given particle, without having to worry about 
calculating different Jacobians.  There will be a dependence on the cutoff 
parameters in both the $2\rightarrow2$ contributions (which include the 
collinear and soft regions) and the $2\rightarrow3$ contributions, but this 
dependence will disappear once the two contributions are added together.

One final point needs to be addressed concerning the subprocess $q \overline q 
\rightarrow \gamma Q \overline Q$. There is a collinear singularity 
associated with the region where the final state $Q \mbox{\rm \ and }\overline 
Q$ are collinear. Physically, this corresponds to a $\gamma g$ final state 
with the gluon splitting into the $Q\overline Q$ pair. Normally, this singular 
region would be integrated over yielding a two-body contribution dependent on 
$\delta_c$ which would be proportional to the subprocess $q \overline q 
\rightarrow \gamma g$. This would be added to the one-loop corrections 
for the $q \overline q \rightarrow \gamma g$ subprocess, the poles in 
$\epsilon$ would cancel and there would be a residual $\delta_c$ contribution 
to the $q \overline q \rightarrow \gamma g$ subprocess. This would cancel 
against a similar contribution from $q \overline q \rightarrow \gamma Q 
\overline Q$ once a suitable jet definition was implemented in the 
calculation. However, once one tags the jet as containing a heavy quark, the 
problem arises in that there is no contribution from the subprocess 
$q \overline q \rightarrow \gamma g$. Hence, there is an uncanceled 
$\delta_c$ dependence in the $2 \rightarrow 3$ contribution. This problem
is addressed by realizing that physically the final state gluon can not 
fragment into a $Q \overline Q$ pair unless its invariant mass exceeds 
$4m_Q^2$. Imposing this constraint on the events generated for $q \overline q 
\rightarrow \gamma Q \overline Q$ avoids the problem of the uncanceled 
$\delta_c$ dependence.
  
\section{Results}
\subsection{Tevatron Predictions}   
For the numerical results shown below the CTEQ6.6M PDFs \cite{Cteq:66M} were 
used, unless otherwise stated, with a 2-loop $\alpha_s$ corresponding to  
$\alpha_s(M_Z)=0.118$. The cross section was calculated for a center of mass 
energy of $\sqrt{S}=1.96{\rm\ TeV}$ corresponding to the measurements 
being made at the Tevatron. The cuts applied reproduce the ones used by the 
D\O\  experiment, where the lower bounds for the transverse momenta of the 
photon and heavy quark are as follows: $p_{T \gamma}>30 {\rm\ GeV}, 
p_{TQ}>15 {\rm\ GeV}$, and their rapidities are limited to the central region 
of the detector $|y_{\gamma}|<1,|y_b|<0.8$.  If two final state partons lie 
within a cone of radius $\Delta R=0.5$, where 
$R=\sqrt{\Delta \eta^2 +\Delta \phi^2}$ then they are merged into a single 
jet.  If a final state has two heavy quark jets within the detectable region, 
it is counted only once, taking into account the transverse momentum of the 
more energetic jet.  To be experimentally detectable, a photon needs to be 
isolated. This means that it should not be surrounded by hadronic energy more 
then $E_h=\epsilon*E_{\gamma}$ in a cone of radius $R=R_{iso}$ around it.  
The photon isolation requirements imposed model those needed in the D0 
detection of a direct photon and are: $R_1<0.2$, $\epsilon_1<0.04$ and 
$R_2<0.4$, $\epsilon_2<0.07$.  

\begin{figure}
\begin{minipage}[t]{8cm}
\begin{center}
\includegraphics[scale=0.3,angle=270]{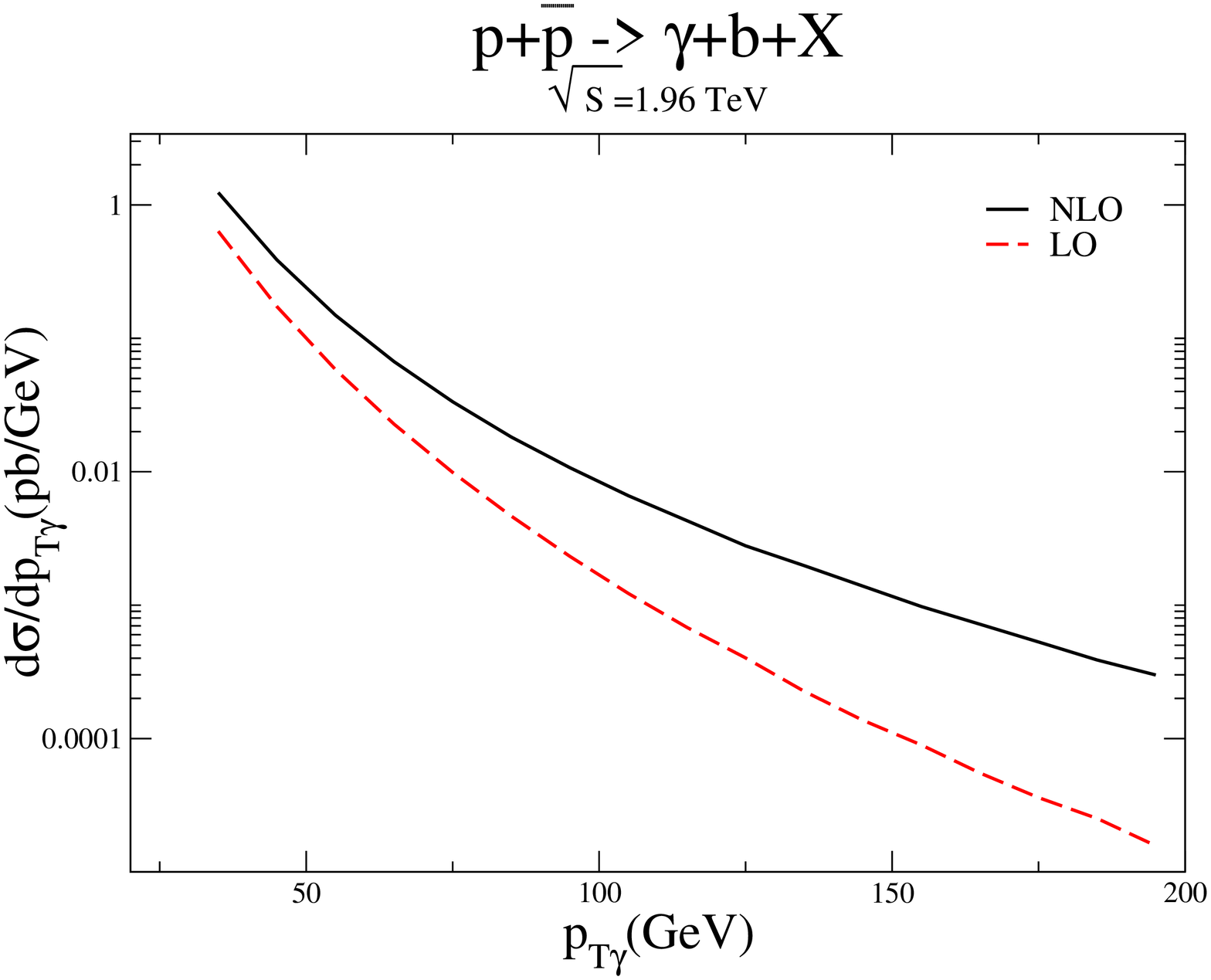}
\caption{\label{NLO_b} \small{The differential cross section, 
$d\sigma /dp_{T\gamma}$ for the production of a direct photon and a bottom 
quark as a function of $p_{T\gamma}$ for $\sqrt{S}=1.96{\rm\ TeV}$, at NLO - 
solid line, and at LO - dashed line }}
\end{center}
\end{minipage}
\hfill
\begin{minipage}[t]{8cm}
\begin{center}
\includegraphics[scale=0.3,angle=270]{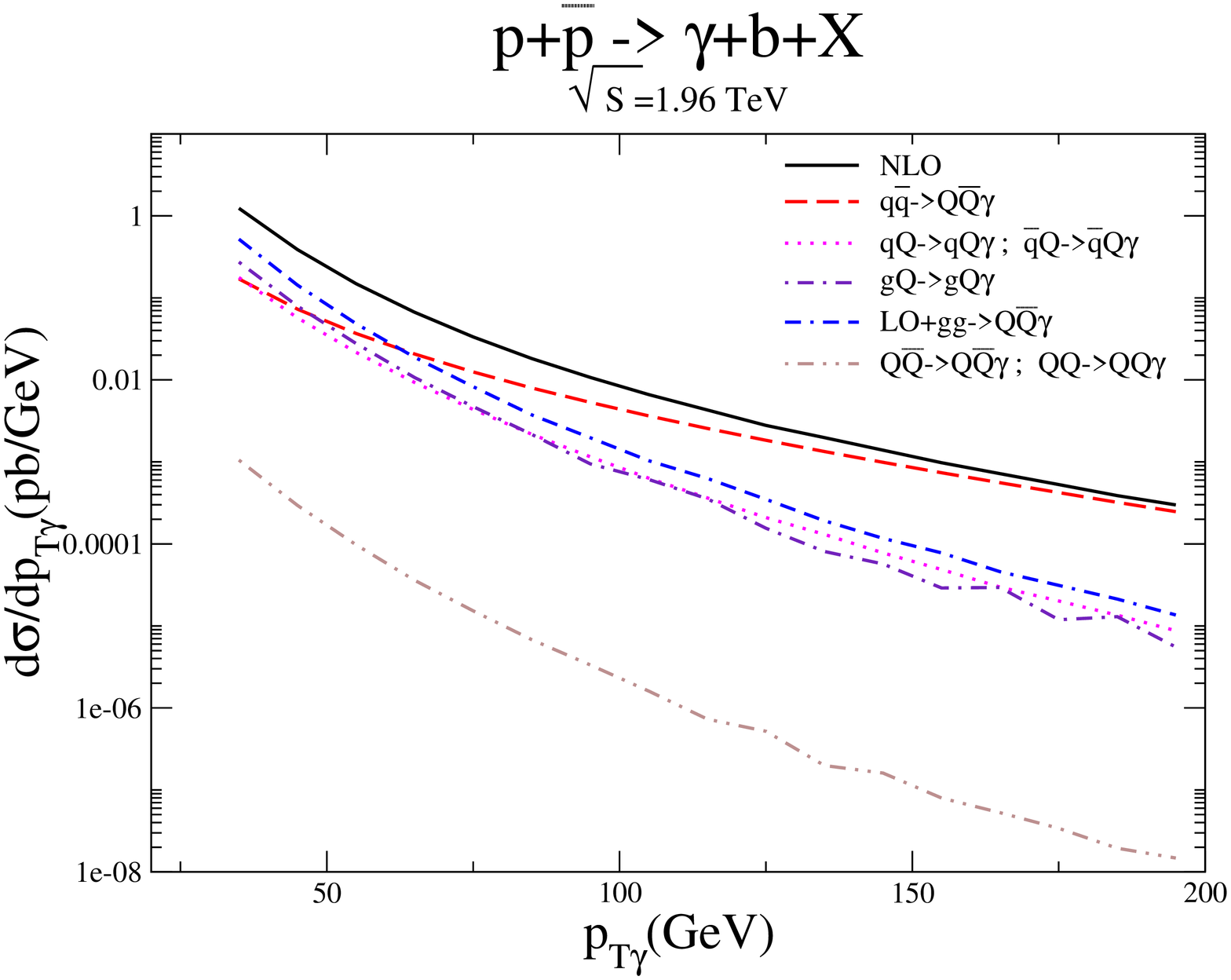}
\caption{\label{parts} \small{contributions of the different subprocesses to 
the differential cross section, NLO -solid line, annihilation 
$q\bar q\rightarrow Q\bar Q\gamma$ - dashed line, 
$qQ\rightarrow qQ \gamma$ - dotted line, $gQ\rightarrow gQ\gamma $ - 
dot dashed line, $gg\rightarrow Q\bar Q\gamma + $LO - dash dot dotted line, 
$Q\bar Q\rightarrow \gamma Q\bar Q$, and $QQ\rightarrow \gamma QQ$ - dot dash 
dotted line}}
\end{center}
\end{minipage}
\end{figure}

The differential cross section for the process $p\bar p\rightarrow \gamma b X$ 
as a function of the transverse momentum of the photon is shown in 
Fig.\ref{NLO_b}. It is interesting to note in Fig.\ref{NLO_b} that as 
$p_{T\gamma}$ grows, the difference between the LO and NLO curves increases 
substantially.  
To understand the origin of this effect it is necessary to look at how 
the different subprocesses listed in Table 1 contribute to the cross section. 
This decomposition is shown in Fig.\ref{parts}.  

It is apparent from Fig.\ref{parts} that the effect shown in Fig.\ref{NLO_b} is 
driven by the annihilation subprocess, $q\bar q\rightarrow \gamma Q\bar Q$, 
which overtakes the Compton contribution and starts dominating the cross 
section at 
$p_{T\gamma}\sim 70 {\rm\ GeV}$.  The $gQ\rightarrow \gamma gQ$ and 
$Qq\rightarrow \gamma qQ$ /$Q\bar q\rightarrow \gamma qQ$ subprocesses 
contribute to the cross section about equally, with $gQ\rightarrow \gamma gQ$ 
prevailing over $Qq\rightarrow \gamma qQ$ /$Q\bar q\rightarrow \gamma qQ$ at 
small $p_{T\gamma}$, where the gluon PDF is larger than the light quark PDF, 
and then at large $p_{T\gamma}$, 
$Qq\rightarrow \gamma qQ$ /$Q\bar q\rightarrow \gamma qQ$ takes over when 
the light quark PDFs become larger than the gluon PDFs.  The Compton 
subprocess and $gg\rightarrow \gamma Q\bar Q$ are added together, since the 
$gg\rightarrow \gamma Q\bar Q$ contribution is negative. This negative 
contribution is what remains after the appropriate collinear terms are 
subtracted. The size of the $2\rightarrow3 \ gg $ contribution is scale 
dependent, with the compensating collinear terms contributing to the 
$2\rightarrow2$ component. The role of the 
$Q\bar Q\rightarrow \gamma Q\bar Q$ / $QQ\rightarrow \gamma QQ$ subprocesses 
is almost negligible, as the heavy quark PDFs are much smaller than the light 
quark and gluon PDFs. 

\begin{figure}[h]
\begin{center}
\includegraphics[scale=1.0,angle=0]{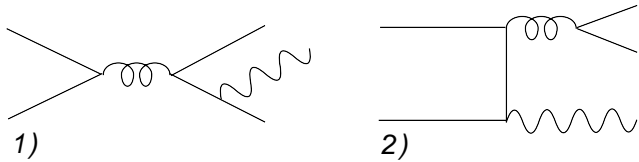}
\end{center}
\caption{\label{qqb} \small{Some typical Feynman diagrams for the 
annihilation subprocess 
$q\bar q\rightarrow \gamma Q\bar Q$ where 1) the photon is emitted from the 
final 
state heavy quarks and 2) the photon is emitted from the initial state light 
quarks}}
\end{figure}

\begin{figure}
\begin{minipage}[t]{8cm}
\begin{center}
\includegraphics[scale=0.3,angle=270]{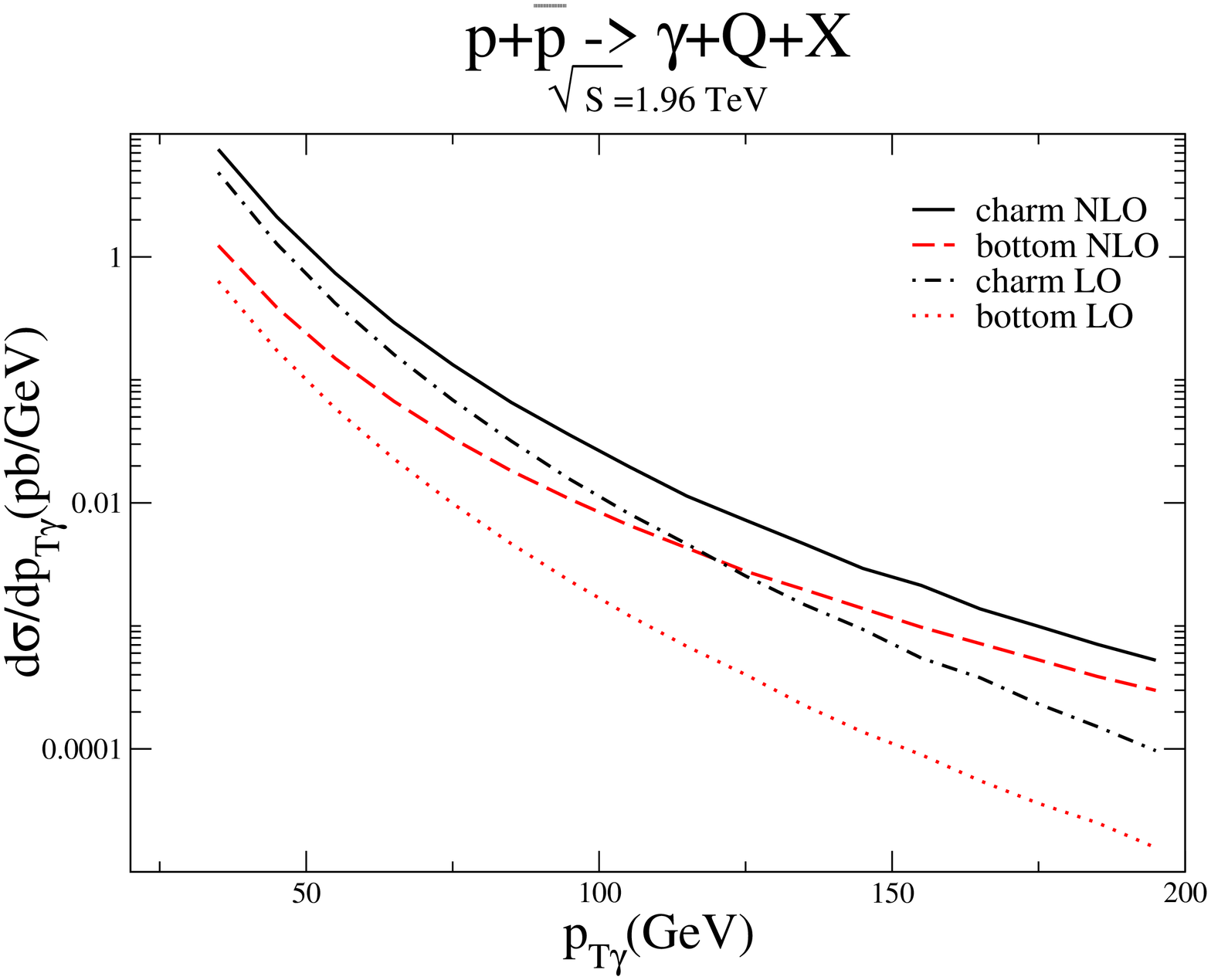}
\caption{\label{b_c} \small {a comparison between the differential cross 
sections, $d\sigma /dp_{T\gamma}$ for the production of a direct photon and a 
bottom quark, and that of a direct photon plus a charm quark at NLO and LO,  
charm at NLO - solid line and for bottom at NLO- dashed line, charm at LO - 
dot dashed line, bottom at LO - dotted line}}
\end{center}
\end{minipage}
\hfill
\begin{minipage}[t]{8cm}
\begin{center}
\includegraphics[scale=0.3,angle=270]{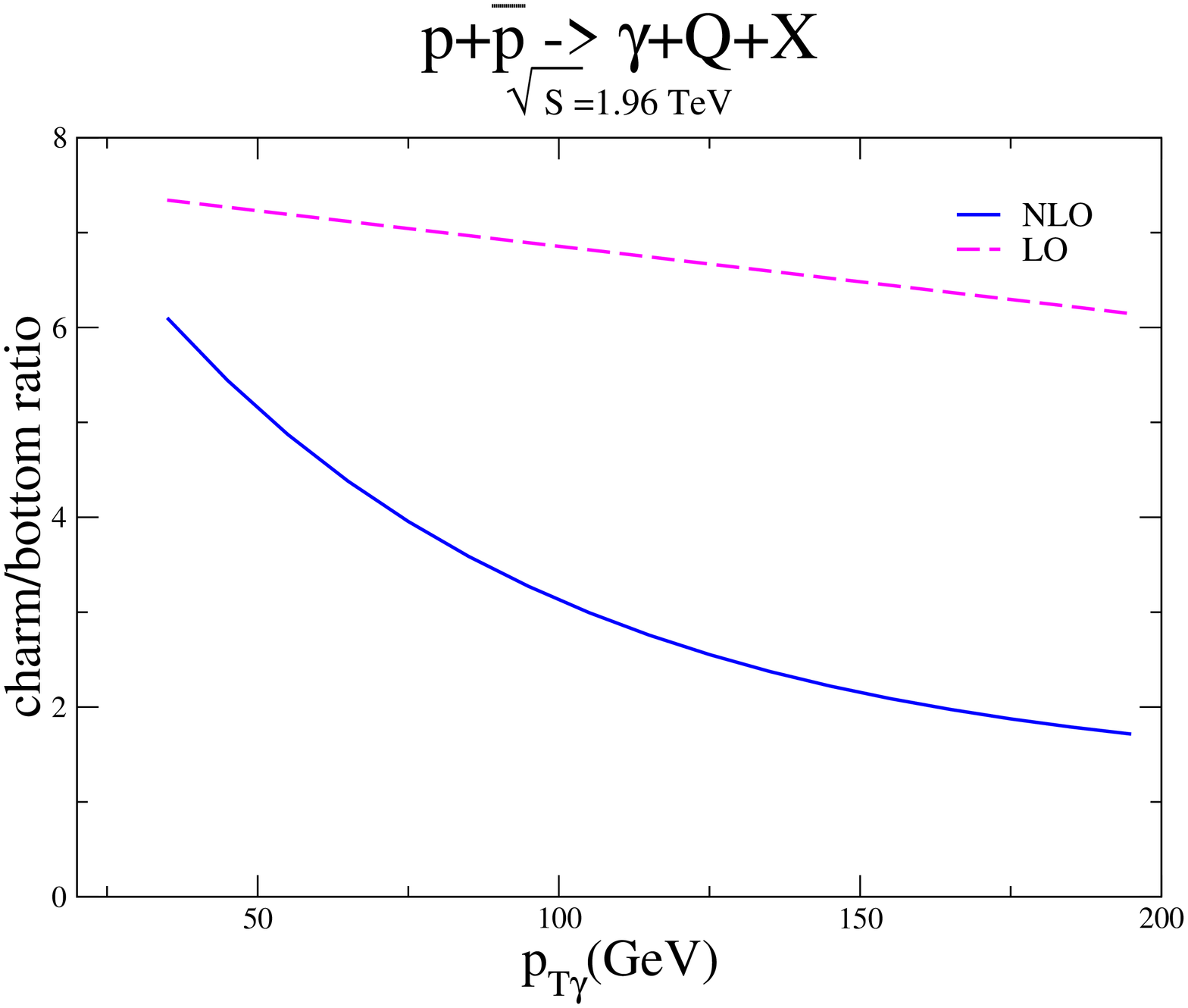}
\caption{\label{b_c-ratio} \small{the ratio of the charm and bottom 
differential cross sections versus $p_{T\gamma}$, at NLO - solid line and at 
LO - dashed line}}
\end{center}
\end{minipage}
\end{figure}

Since the annihilation subprocess dominates 
the cross section at large $p_{T\gamma}$, it is useful to look at some of 
the Feynman diagrams contributing to it, as shown in Fig.\ref{qqb}. 
There are two channels through which the annihilation subprocess can be 
produced, an s-channel shown in diagram 1) and a t-channel in diagram 2) of 
Fig.\ref{qqb}.  Since the photon couples to the final state heavy quarks in 
diagram 
1), this diagram is proportional to the heavy quark charge, 
$e_Q$, whereas 
in diagram 2) the photon couples to the initial state light quarks, and thus 
this diagram is proportional to the light quark's charge, 
$e_q$.  Diagram 2) begins to 
dominate as $p_{T\gamma}$ grows and, since it does not depend on the heavy quark 
charge, we expect the difference between the bottom and charm cross sections 
to decrease as $p_{T\gamma}$ increases.  This indeed is the case as can be seen 
from Fig.\ref{b_c}, where the NLO differential cross section for the charm 
quark (solid line) and the one for the bottom quark (dashed line) tend to come 
closer as the value of the transverse momentum increases.  However, the 
difference 
between the LO cross sections stays about the same as can be seen from 
the dot-dashed (charm) and dotted curves (bottom) in Fig.\ref{b_c}, and also 
from Fig.\ref{b_c-ratio}, where the ratio of the NLO and LO charm and bottom 
differential cross sections is shown.  The ratio of the LO cross sections 
stays almost constant since the main contribution to the LO cross section 
comes from the Compton subprocess, with the difference between the charm and 
bottom curves arising from the difference in the charges of the charm and 
bottom quarks and the relative sizes of the heavy quark PDFs.  The ratio of 
the two LO cross sections depends on the ratio of 
the charges squared which is $e_c^2/e_b^2=4$, and is driven up from that value 
to about $\sim 7$ because the charm PDF is larger than the bottom PDF.  

The fact that the annihilation subprocess dominates the cross section at large 
$p_{T\gamma}$ also increases the scale dependence of the cross-section in that 
region. There is no Born term which involves a $q \bar q$ initial state 
and, therefore, the contributions from the annihilation subprocess 
start in \cal{O}$(\alpha \alpha_s^2)$. As such, the typical compensation 
between LO and NLO contributions for this subprocess is missing and the 
annihilation subprocess can be thought of a leading order.


\begin{figure}[t]
\begin{center}
\includegraphics[scale=0.3,angle=270]{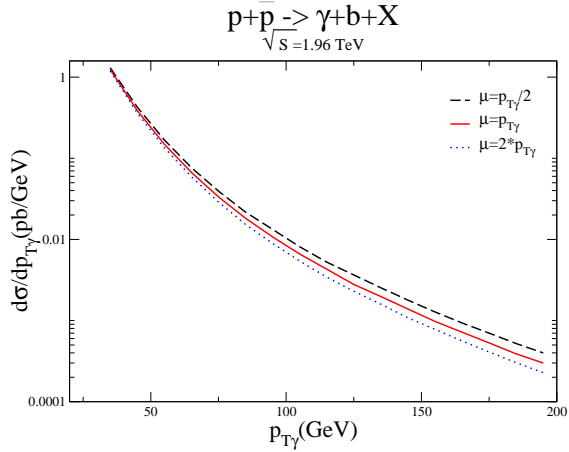}
\end{center}
\caption{\label{scale} \small{Scale dependence of the differential cross 
section, $d\sigma /dp_{T\gamma}$ for the production of a direct photon and a 
bottom quark, where the three different scales have been set to be equal 
$\mu=\mu_r=\mu_f=\mu_F$, $\mu=p_{T\gamma}$ - solid line, $\mu=p_{T\gamma}/2$ - 
dashed line, $\mu=2p_{T\gamma}$ - dotted line }}
\end{figure}

As can be seen from Fig.\ref{scale} the scale dependence increases at large 
$p_{T\gamma}$, where the annihilation starts to dominate.  The renormalization, 
$\mu_r$, factorization, $\mu_f$ and fragmentation, $\mu_F$ scales have been 
set to be equal and are denoted by $\mu$.

\begin{figure}
\begin{minipage}[t]{8cm}
\begin{center}
\includegraphics[scale=0.3,angle=270]{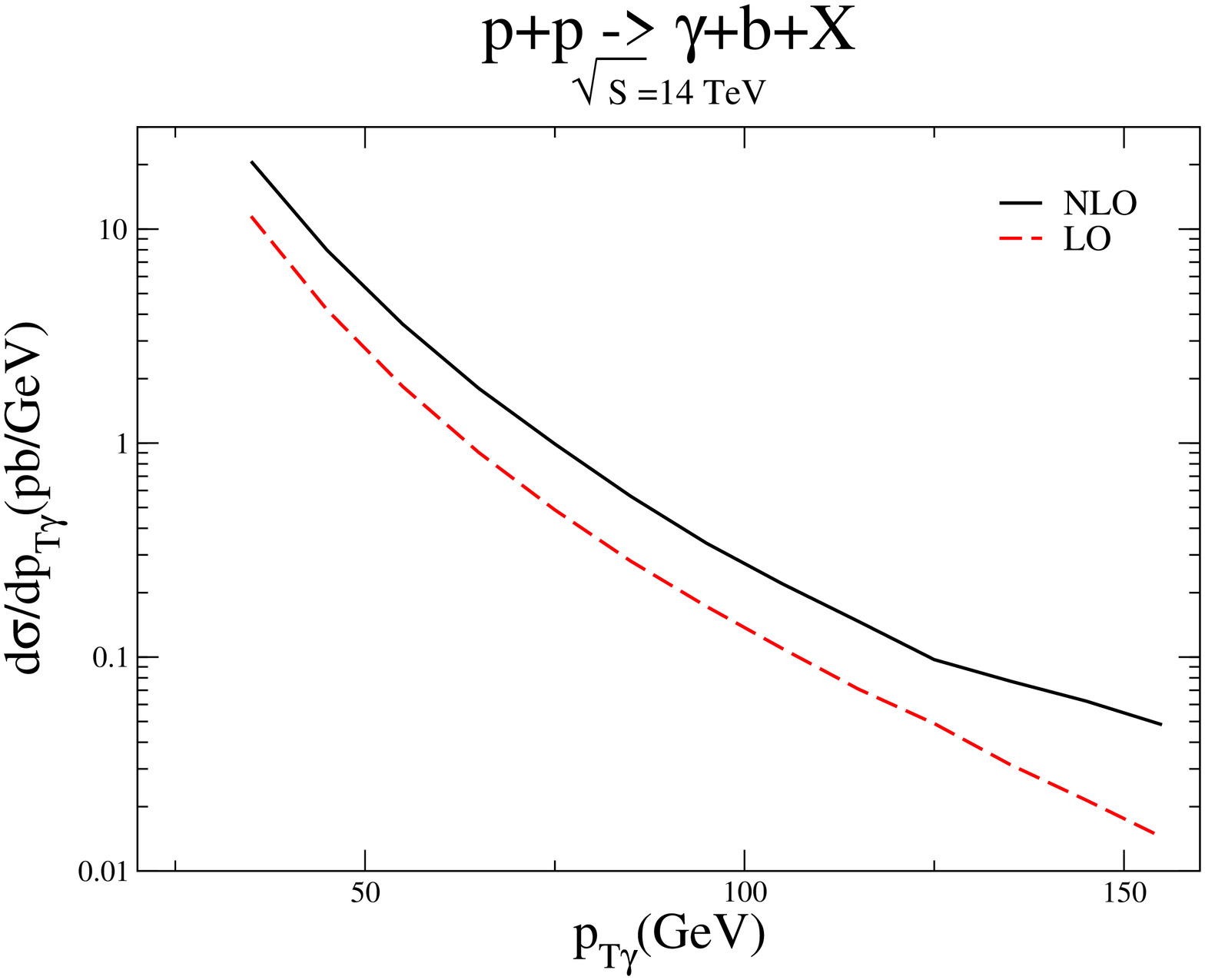}
\caption{\label{LHC} \small{the differential cross section versus the 
transverse momentum of the photon $d\sigma /dp_{T\gamma}$ for the production of 
a direct photon and a bottom quark at LHC center of mass energies, 
$\sqrt{S}=14{\rm\ TeV}$, NLO - solid line, LO - dashed line}}
\end{center}
\end{minipage}
\hfill
\begin{minipage}[t]{8cm}
\begin{center}
\includegraphics[scale=0.3,angle=270]{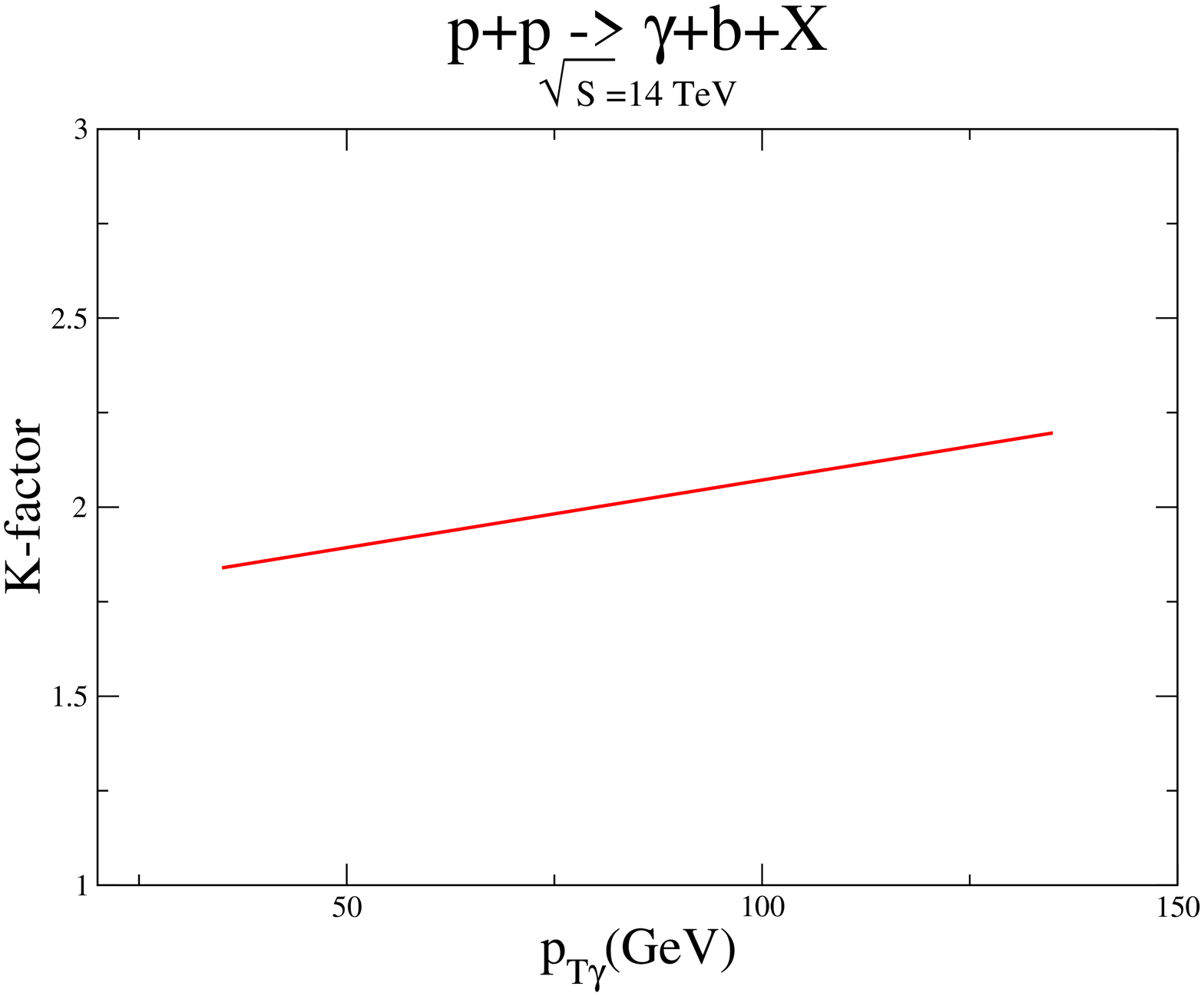}
\caption{\label{Kfac} K factor, or the ratio of the NLO to the LO differential 
cross section  for $pp\rightarrow b \gamma X$  at $\sqrt{S}=14{\rm\ TeV}$}
\end{center}
\end{minipage}
\end{figure}

\subsection{LHC Predictions}

\begin{figure}[t]
\begin{center}
\includegraphics[scale=0.3,angle=270]{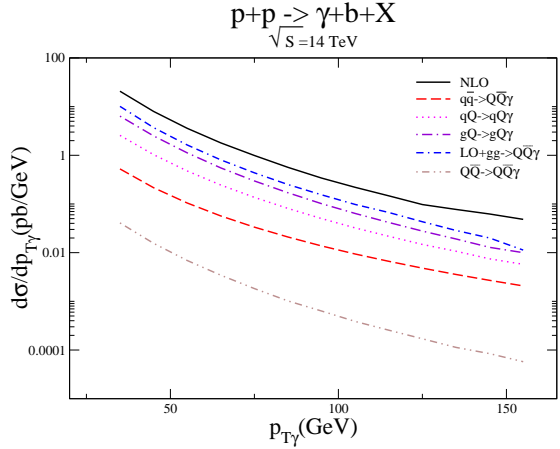}
\end{center}
\caption{\label{LHC-p} \small{contributions of the different subprocesses to 
the differential cross section, $d\sigma /dp_{T\gamma}$ for $pp\rightarrow b 
\gamma X$  at $\sqrt{S}=14{\rm\ TeV}$, NLO -solid line, annihilation $q\bar 
q\rightarrow Q\bar Q\gamma$ - dashed line, $qQ\rightarrow qQ \gamma$ - dotted 
line, $gQ\rightarrow gQ\gamma $ - dot dashed line, $gg\rightarrow Q\bar 
Q\gamma + $LO - dash dot dotted line, $Q\bar Q\rightarrow \gamma Q\bar Q$, 
and $QQ\rightarrow \gamma QQ$ - dot dash dotted line }}
\end{figure}

When data become available from the Large Hadron Collider (LHC), it will be 
very important to have a good understanding of what Standard Model (SM) 
processes are going to look like at center of mass energies of 
$\sqrt{S}=14{\rm\ TeV}$, as these processes will provide important 
means of calibrating and understanding the detectors and, ultimately, are 
likely to provide significant backgrounds to new physics signals. The 
differential cross section versus the transverse 
momentum of the photon is shown in Fig.\ref{LHC}. It is apparent that the 
increase of the difference between the NLO (solid line) and the LO 
(dashed line) grows much less rapidly with 
increasing $p_{T\gamma}$ than was the case for the Tevatron.  In Fig.\ref{Kfac} 
the K-factor, which is the ratio between the NLO and LO cross section for b 
quarks is shown, which 
stays stable and is around $2$.  To understand the difference between the 
LHC and Tevatron curves, the contributions of the different parts contributing 
to the LHC cross section are shown in Fig.\ref{LHC-p}.  
From Fig.\ref{LHC-p} it can be seen that the annihilation subprocess no 
longer drives the cross section at high $p_{T\gamma}$, and now it is the LO, 
and the $gQ\rightarrow \gamma gQ$ subprocesses that are the most prominent.  
These differences come about for two reasons.  As the LHC collides two beams 
of protons, instead of the proton and antiproton beams at Fermilab, there is 
no longer any valence light antiquarks present. Hence, the relative 
contribution of the annihilation subprocess is decreased.  Also, because the 
LHC will ultimately operate at a center of 
mass energy which is about seven times larger than that of the Tevatron, 
lower values of $x \sim p_T/\sqrt{s}$ are probed at the LHC. 
For the kinematic region shown in Fig. \ref{LHC-p} the gluon 
PDF is dominant, accounting for the continued importance of the $gQ$
initiated subprocesses.

An interesting consequence of this pattern of subprocess contributions is 
that the the dominant parts are all proportional to the heavy quark PDFs. 
Such was not the case for the Tevatron curves, except for the low end of the 
$p_{T\gamma}$ range. Accordingly, heavy quark + photon measurements at the LHC 
will have the potential to provide important cross checks on the 
perturbatively calculated heavy quark PDFs. These PDFs are likely to provide 
important contributions to other physics signals -- either standard model 
or new physics -- and such checks will be an important part of the search for 
new physics.

\subsection{NLO Fragmentation and Photon Isolation}

\begin{figure}
\begin{minipage}[t]{8cm}
\begin{center}
\includegraphics[scale=0.3,angle=270]{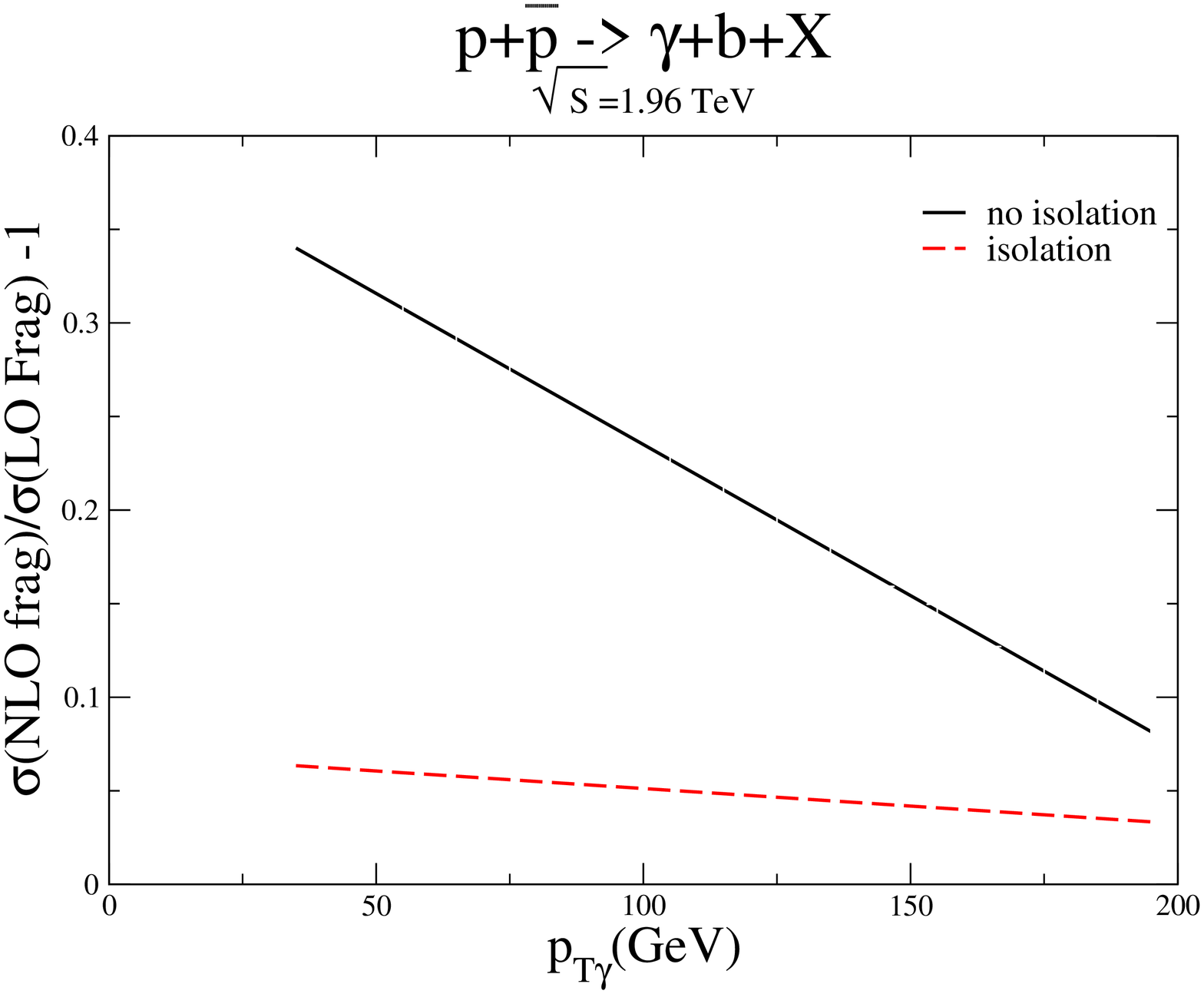}
\caption{\label{ratio_iso} \small{ratio between the differential cross section 
$d\sigma /dp_{T\gamma}$, with NLO fragmentation contribution included and the 
differential cross section with just LO fragmentation included, solid line no 
isolation required, dashed line -isolation}}
\end{center}
\end{minipage}
\hfill
\begin{minipage}[t]{8cm}
\begin{center}
\includegraphics[scale=0.3,angle=270]{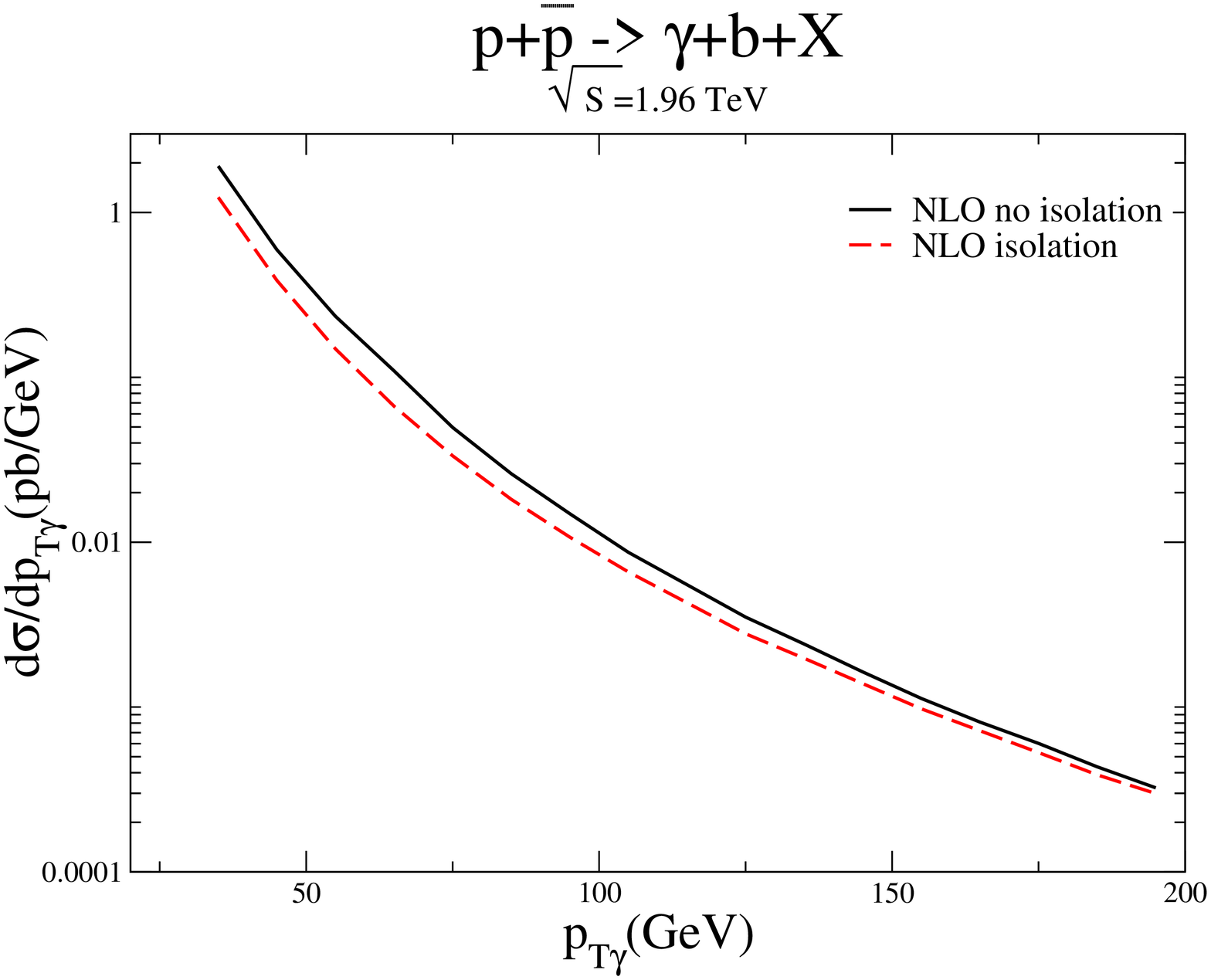}
\caption{\label{iso} \small{comparison between the differential cross section, 
$d\sigma /dp_{T\gamma}$ without isolation requirements and with them, no 
isolation - solid line , isolation - dashed line}}
\end{center}
\end{minipage}
\end{figure}

It is interesting to investigate what effect the NLO Fragmentation 
contributions have upon the cross section.  Fig.\ref{ratio_iso} shows the ratio 
between the full NLO calculation and the cross section with only LO 
fragmentation. If there are no isolation 
requirements imposed on the photon, the cross section increases up to 
$\sim 30\%$, solid curve in Fig.\ref{ratio_iso}.  As mentioned above a photon 
needs to be isolated in order to give a clear signal at a detector.  The 
isolation requirements affect the photon which is produced by fragmentation 
the strongest, as it is emitted in close proximity to the parton from which 
it is fragmenting.  This can be seen from the dashed line in Fig.
\ref{ratio_iso}, where the NLO fragmentation contribution has now decreased to 
a few $\%$.  Fig.\ref{iso} shows the comparison between the differential cross 
section with the inclusion of isolation and without it.  As can be seen this 
difference is larger at low $p_{T\gamma}$, but the two curves come close to 
each other with the increase of the photon's transverse momentum, where as 
seen from Fig.\ref{parts} the $q\bar q\rightarrow \gamma Q\bar Q$ subprocess 
takes over the cross section.

\subsection{Intrinsic Charm}

\begin{figure}
\begin{minipage}[t]{8cm}
\begin{center}
\includegraphics[scale=0.3,angle=270]{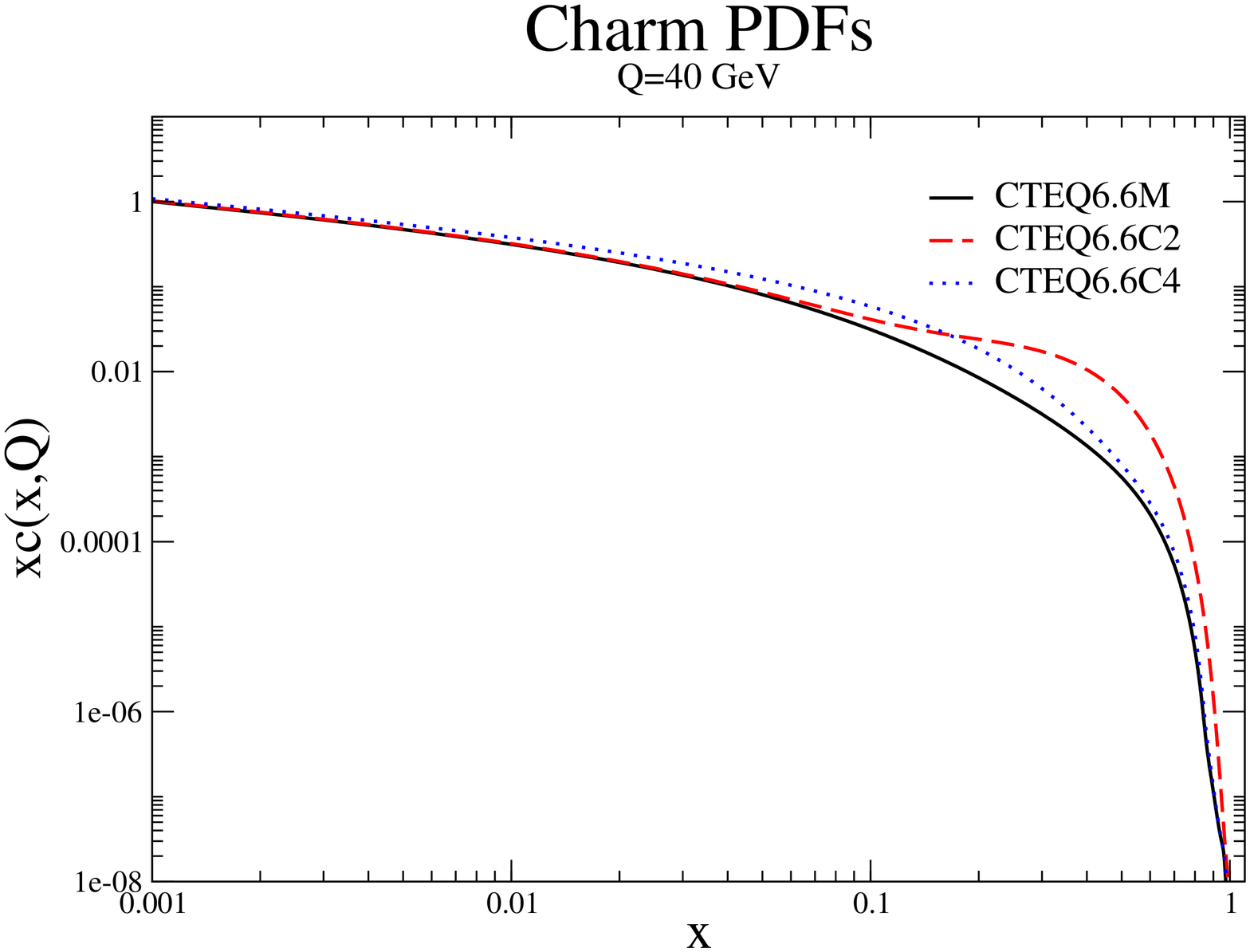}
\caption{\label{ic} \small{comparison between the three different charm PDFs 
at scale $Q=40 {\rm\ GeV}$, CTEQ6.6M - solid line, CTEQ6.6C2 - dashed line, 
CTEQ6.6C4 - dotted line}}
\end{center}
\end{minipage}
\hfill
\begin{minipage}[t]{8cm}
\begin{center}
\includegraphics[scale=0.3,angle=270]{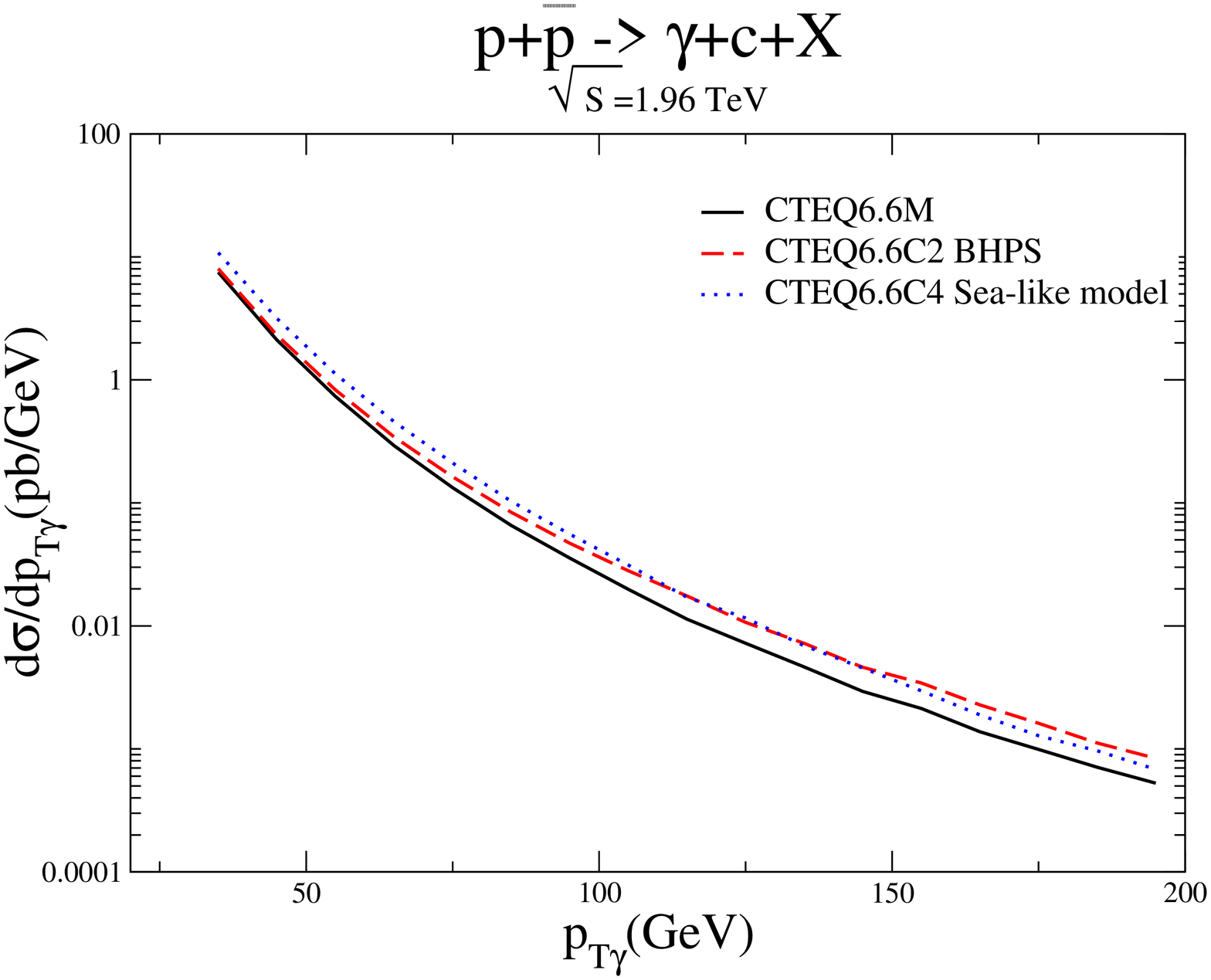}
\caption{\label{iso_nloff} \small{the differential cross section, 
$d\sigma /dp_{T\gamma}$, for the production of a direct photon and a charm 
quark for the three different PDF cases, CTEQ6.6M - solid line, CTEQ6.6C2 - 
dashed line, CTEQ6.6C4 - dotted line}}
\end{center}
\end{minipage}
\end{figure}

In the CTEQ6.6M PDFs used in the previous sections, the charm quark is 
radiatively generated from the gluon's PDF with the use of the DGLAP 
equations.  Thus it follows that there are no charm quarks present at scales 
below the charm mass,$m_c$, or that the charm PDF, $c(x,\mu)=0$, when 
$\mu<m_c$.  This however does not need to be the case, and there are models 
that study the possibility for an intrinsic charm component of the nucleon 
\cite{Cteq:IC}. Two such models are the BHPS model, which is a light-cone 
model, and the sea-like model in which the charm distribution follows the 
shape of the light flavor sea quarks.  The difference between the three cases 
is shown in Fig.\ref{ic}, where the solid curve shows the CTEQ6.6M or 
radiatively generated charm scenario, the dashed curve is the CTEQ6.6C2, or 
BHPS model, and the dotted curve is CTEQ6.6C4 PDF or the sea-like model.  The 
difference between the BHPS distribution and the radiatively generated case 
are most noticeable at large x, whereas the sea-like model is about equally 
larger than the CTEQ6.6M PDF at all values of x.  How these different PDFs 
affect the cross section can be seen from Fig.\ref{iso_nloff}. The dotted 
curve shows the cross section generated with the use of the sea-like intrinsic 
charm PDF, and it is larger than the solid curve by about the same amount 
at all values 
of $p_{T\gamma}$.  The difference between the radiatively generated charm cross 
section and the BHPS charm however is not great at small transverse momentum, 
but it increases at large $p_{T\gamma}$, as is expected given the differences 
between the CTEQ6.6M and CTEQ6.6C2 PDFs at large x.     

\section{Conclusion}  

We have presented the results for the inclusive cross section for the 
production of a direct photon in association with a heavy quark, 
$p \bar p / pp \rightarrow \gamma Q X$, up to $O(\alpha \alpha_s^2)$ with NLO 
fragmentation included.  The inclusion of NLO 
fragmentation has a noticeable effect on the cross section if no isolation is 
imposed.  However, this effect decreases if isolation cuts,
needed for a clean photon signal, are imposed.  
Predictions were presented for $p\bar p$ collisions at $\sqrt{S}=1.96 TeV$ and 
for $pp$ collisions at 14 TeV.
At the Tevatron, due to the $p \bar p$ beams, the valence quarks and 
antiquarks are dominant, and thus it is the annihilation subprocess 
$q\bar q\rightarrow \gamma Q\bar Q$ that dominates the cross section at large 
$p_{T\gamma}$.  Therefore the sensitivity to the initial state heavy quarks and 
their content in the nucleon decreases, and the difference between the bottom 
and charm differential cross sections, $d\sigma /dp_{T\gamma}$, also 
diminishes.  At the LHC, where two beams of protons are colliding, there are 
no longer any valence antiquarks present, and processes with initial gluons 
and heavy quarks dominate.  Thus there should be a greater possibility to 
learn more about the heavy quark role in the nucleon. In particular, the 
perturbatively calculated heavy quark PDFs may be checked using such data. 


\begin{thebibliography}{99}

\bibitem{JO:review}
J.F. Owens, Rev. Mod. Phys. {\bf 59},465 (1987).

\bibitem{ABFOW}
P. Aurenche {\it et al.,} Phys. Rev. D {\bf 39}, 3275 (1989).

\bibitem{Huston:gamma}
J. Huston {\it et al.,} Phys. Rev. D {\bf 51}, 6139 (1995).

\bibitem{Aurenche:old_study}
P. Aurenche {\it et al.,} Eur. Phys. J. C {\bf 9}, 107 (1999).

\bibitem{Aurenche:new_study}
P. Aurenche {\it et al.,} Phys. Rev. D {\bf 73}, 094007 (2006).

\bibitem{Dzero:photonjet}
D0 Collaboration: V. Abazov, {\it et al,.} arXiv:0804.1107v2 [hep-ex], 2008.

\bibitem{Bailey:charm}
B. Bailey, E.L. Berger, and L.E. Gordon, Phys.Rev. D {\bf 54}, 1896 (1996).

\bibitem{Berger:analyt}
E.L. Berger and L.E. Gordon, Phys.Rev. D {\bf 54}, 2279 (1996).

\bibitem{Vogel:mass}
M. Stratmann and W. Vogelsang, Phys.Rev. D {\bf 52}, 1535 (1995).

\bibitem{Berger:spin}
E.L. Berger and L.E. Gordon Phys.Rev. D {\bf 58}, 114024 (1998).

\bibitem{Owens:PS}
B.W. Harris and J.F. Owens, Phys.Rev. D {\bf 65}, 094032 (2002).

\bibitem{Cteq:66M}
P.M. Nadolsky, Phys.Rev. D {\bf 78}, 013004 (2008).

\bibitem{Cteq:IC}
J. Pumplin, H.L. Lai, and W.K. Tung, Phys.Rev. D {\bf 75}, 054029 (2007).

\end{thebibliography}

\end{document}